\documentstyle[aps]{revtex}
\begin{document}
\draft
\title{Current-carrying ground states in mesoscopic and macroscopic systems}
\author{Michael R. Geller}
\address{Institute for Theoretical Physics, University of California,
Santa Barbara, California 93106}
\address{and Department of Physics, University of Missouri,
Columbia, Missouri 65211}
\date{\today}
\maketitle

\begin{abstract}
We extend a theorem of Bloch, which concerns the net orbital current carried by
an interacting electron system in equilibrium, to include mesoscopic effects.
We obtain a rigorous upper bound to the allowed ground-state current in a
ring or disc, for
an interacting electron system in the presence of static but otherwise
arbitrary
electric and magnetic fields. We also investigate the effects of spin-orbit
and current-current interactions on the upper bound. Current-current
interactions,
caused by the magnetic field produced at a
point ${\bf r}$ by a moving electron at ${\bf r}',$
are found to reduce the upper bound by an amount that is determined by the
self-inductance of the system.
A solvable model of an electron system
that includes current-current interactions is shown to realize our upper bound,
and the upper bound is compared with measurements of the persistent current in
a single ring.
\end{abstract}


\section{INTRODUCTION}

A theorem due to Bloch holds that an interacting electron system in equilibrium
carries
{\it no} net orbital current \cite{Bohm}.
This question originally had been motivated by early attempts, before the
Bardeen-Cooper-Schrieffer theory, to explain superconductivity by proposing
that
electron-electron interactions lead to special current-carrying states of lower
energy than the current-free states. However, it is now understood that
supercurrent-carrying
states are in fact metastable {\it nonequilibrium} states, which,
because of their off-diagonal long-range
order or wave function rigidity, have an extremely long lifetime.

The great interest over the past several years in the physics of mesoscopic
systems again has
made the question of allowed equilibrium currents an important one.
More than ten years ago,
B\"uttiker {\it et al.} \cite{Buttiker etal} predicted the existence of
equilibrium currents in mesoscopic normal-metal rings threaded by a magnetic
flux.
Recent experimental evidence \cite{Levy etal,Chandresakhar etal} in support of
this conjecture has
stimulated considerable interest in these so-called persistent currents.
Although a satisfactory
explaination of the experiments is still lacking, the present consensus is that
both
electron-electron interaction and disorder effects are important
\cite{Ambegaokar and Eckern}.
A related phenomena is that of {\it spontaneous} orbital
currents occurring in the absence
of any applied magnetic field or twisted boundary conditions.
Although there is no experimental
evidence for this symmetry-breaking state, spontaneous
orbital currents have been predicted to
occur by several authors \cite{Wohlleben etal,Rasolt and Perrot,Choi}.

Given the diverse situations in which equilibrium current-carrying
states may occur, it is
worthwhile to revise Bloch's theorem to incorporate these
mesoscopic effects. To this end,
Vignale \cite{Vignale} has recently derived a rigorous upper bound to the
persistent
current in a single ring, the results being valid for both
noninteracting and interacting
electrons in the presence of arbitrary magnetic fields and
impurity potentials. One surprising result
of Vignale's analysis is that although the upper bound
to the persistent current in a thin ring
of uniform density and radius $R$ vanishes as $1/R$ for large $R$, it does {\it
not} vanish
for a thick ring or punctured disc with the ratio $R_{\rm in}/R_{\rm out}$ of
the inner
radius and outer radius fixed as these radii become infinite. Although
Vignale's result does not
preclude the existence of a more stringent upper bound that
always vanishes in the macroscopic limit,
the upper bound is actually realized in calculations
of the persistent current (the integrated
azimuthal current density) in a two-dimensional noninteracting
electron gas in a large quantum
dot \cite{Avishai and Kohmoto,Geller and Vignale}.

One motivation for this work is to extend the analysis of Ref.~\cite{Vignale}
to
include the effects of spin-orbit interaction,
which has received considerable attention in connection with persistent
currents
and spontaneous currents. Spin-orbit interaction is known to lead to a
topological
interference effect, called the Aharanov-Casher effect \cite{Aharanov and
Casher},
which is an electromagnetic dual of the Aharanov-Bohm effect. Meir {\it et al.}
\cite{Meir etal} have shown that spin-orbit scattering in one-dimensional
disordered rings
induces an effective magnetic flux which reduces the persistent current in a
universal manner.
The effect of spin-orbit interaction on mesoscopic persistent currents has been
studied by
several other authors, who also find reduced currents
\cite{Loss etal,Balatsky and Altschuler,Fujimoto and Kawakami,Qian and Su}.
This has led us to question whether the upper bound on the allowed persistent
current
is itself reduced by spin-orbit interactions. We shall show here that this is
not the case.

A second motivation for this work is to examine the infleuence
of current-current interactions,
an order $v^2/c^2$ relativistic effect caused by the magnetic
field produced at a point ${\bf r}$
by a moving electron at ${\bf r}'$. The possibility of these magnetic
interactions leading to a
spontaneous current-carrying state in mesoscopic metal rings has been discussed
in a remarkable
paper by Wohlleben {\it et al.} \cite{Wohlleben etal},
where it is shown that, in zero field, a small
ring exhibits a transition to a state with persistent current. The combined
effects of
spin-orbit coupling and current-current interactions has been studied recently
by Choi
\cite{Choi}. We shall show below that current-current interactions reduce
the upper bound on the ground-state
current by an amount which is determined by the self-inductance of the system.

In this paper, we derive an upper bound to the ground-state current in an
arbitrary
many-electron system, including spin-orbit coupling and current-current
interaction
effects. To best demonstrate the modifications to Bloch's theorem from the
effects
of finite sample size, we restrict our analysis to zero temperature. However,
our final results
are also valid at finite temperature, as may be shown by following the method
of
Ref. \cite{Vignale}.

\section{rigorous upper bound}

We begin by obtaining a many-electron Hamiltonian that includes current-current
interactions
to order $v^2/c^2$.
In the transverse gauge, the vector potential seen by an electron at ${\bf
r}_n$ in the
presence of the other moving electrons (of charge $-e$) is
\begin{equation}
A^i({\bf r}_n) = - {e \over 2c} \sum_{n' \neq n}
{T^{ij}({\bf r}_n - {\bf r}_{n'}) \  v_{n'}^j \over
| {\bf r}_n - {\bf r}_{n'}| } ,
\end{equation}
where ${\bf v}_n$ is the velocity of the $n$th electron, and where
\begin{equation}
T^{ij}({\bf r}) \equiv \delta^{ij} + { r^i
r^j \over |{\bf r}|^2 }.
\end{equation}
This vector potential leads to a current-current
interaction term in the Lagrangian of the form
\begin{equation}
L_{\rm int} = {e^2 \over 2 c^2} \sum_{n < n'} { v_n^i \ T^{ij}({\bf r}_n -
{\bf r}_{n'}) \ v_{n'}^j \over
|{\bf r}_n - {\bf r}_{n'}|},
\end{equation}
which in turn leads to a current-current interaction term in the Hamiltonian of
the form
\begin{equation}
H_{\rm int} = - {e^2 \over 2 m^2 c^2} \sum_{n < n'} p_n^i { T^{ij}({\bf r}_n -
{\bf r}_{n'}) \over
| {\bf r}_n - {\bf r}_{n'} | } p_{n'}^j,
\end{equation}
to leading order in $v^2/c^2$. The complete Hamiltonian, including spin-orbit
coupling,
Coulomb and current-current interactions, and coupling to additional electric
and magnetic
fields, may be written as
\begin{equation}
H = \sum_n \bigg( {\Pi_n^2 \over 2m} - {\Pi_n^4 \over 8 m^3 c^2}
+ V({\bf r}_n)
+ {1 \over 2 m^2 c^2} {\bf S}_n \cdot
\big[ \nabla V({\bf r}_n) \times {\bf \Pi}_n \big] \bigg)
+ \sum_{n < n'} {e^2 \over |{\bf r}_n - {\bf r}_{n'}| }
- {e^2 \over 2 m^2 c^2} \sum_{n < n'} \Pi_n^i { T^{ij}({\bf r}_n - {\bf
r}_{n'})   \over
|{\bf r}_n - {\bf r}_{n'}| } \Pi_{n'}^j ,
\label{Hamiltonian}
\end{equation}
where ${\bf \Pi}_n \equiv {\bf p}_n + {e\over c} {\bf A}({\bf r}_n)$,
the $S^i$ are spin operators, and where the potentials ${\bf A}$ and
$V$ are time-independent but otherwise arbitrary. Spin-spin interactions and
the coupling
of the spin degrees of freedom to the external magnetic field are not important
here and shall
be ignored.
The velocity operator ${\bf v}_n \equiv [{\bf r}_n,H]/i\hbar$ is given by
\begin{equation}
v^i_n = {\Pi^i_n\over m}\bigg( 1 - { \Pi_n^2 \over 2 m^2 c^2} \bigg)
 + {1 \over 2 m^2 c^2} [{\bf S}_n \times \nabla V({\bf r}_n)]^i
- {e^2 \over 4 m^2 c^2} \sum_{n' \neq n}
\bigg( \Pi^j_{n'} { T^{ij}({\bf r}_n - {\bf r}_{n'})  \over |{\bf r}_n - {\bf
r}_{n'}|}
+ {T^{ij}({\bf r}_n - {\bf r}_{n'})  \over |{\bf r}_n -
{\bf r}_{n'}|} \Pi^j_{n'} \bigg).
\label{velocity}
\end{equation}

We shall consider a system of $N$ electrons confined to a ring or disc,
oriented with
its axis along the $z$ direction, and we write
(\ref{Hamiltonian}) in cylindrical coordinates ${\bf r} = (r,\theta,z)$.
The thickness and
cross-sectional shape of the system is arbitrary.
The many-body ground state $\psi({\bf r}_{\scriptscriptstyle 1}
s_{\scriptscriptstyle 1},
\cdots , {\bf r}_{\scriptscriptstyle N} s_{\scriptscriptstyle N})$ satisfies
\begin{equation}
H \psi = E \psi,
\end{equation}
where $E$ is the ground-state energy.

Suppose that the ground state $\psi$ carries an
orbital persistent current
\begin{equation}
I = - {e \over 4 \pi}
 \bigg\langle \psi \bigg| \sum_n \bigg(
{ {\bf e}_\theta({\bf r}_n) \over r_n} \cdot {\bf v}_n
+ {\bf v}_n \cdot {{\bf e}_\theta({\bf r}_n) \over  r_n } \bigg)
 \bigg| \psi \bigg\rangle,
\end{equation}
where ${\bf e}_{\theta}({\bf r})$ is an azimuthal unit vector at
${\bf r}$.
We may construct a {\it rotating} state $\psi' = U \psi$, where
\begin{equation}
U \equiv \prod_n e^{i \delta \! L  \theta_n / \hbar},
\end{equation}
which is not necessarily an eigenstate of $H$, and which has a mean energy
$E' \equiv \langle \psi' |H| \psi' \rangle $ given by
\begin{eqnarray}
E' &=& E - {2 \pi \over e} I \ \delta \! L
+ {1 \over 2m} \bigg\langle \sum_n {1 \over r_n^2}
\bigg\rangle {\delta \! L}^2
- {1 \over 8 m^3 c^2} \bigg\langle \sum_n \bigg(
\Pi_n^2 {1 \over r_n^2} + {1 \over r_n^2} \Pi_n^2 + 4 {(\Pi_n^{\theta})^2
 \over r_n^2 } \bigg) \bigg\rangle {\delta \! L}^2 \nonumber \\
&-& {e^2 \over 4 m^2 c^2} \bigg\langle \sum_{n \neq n'}   {
e^i_{\theta}({\bf r}_n) \  T^{ij}({\bf r}_n - {\bf r}_{n'}) \
e^j_{\theta}({\bf r}_{n'})
\over r_n  | {\bf r}_n - {\bf r}_{n'} | r_{n'} } \bigg\rangle {\delta \! L}^2
- {1 \over 2 m^3c^2} \bigg\langle \sum_n
 {\Pi_n^{\theta}  \over r_n^3}
\bigg\rangle  {\delta \! L}^3
- {1 \over 8 m^3 c^2} \bigg\langle \sum_n {1 \over r_n^4} \bigg\rangle {\delta
\! L}^4.
\label{mean energy}
\end{eqnarray}
Here ``$\big\langle \ \ \ \big\rangle $'' denotes an expectation value in the
original ground state $\psi$,
$\Pi_n^\theta \equiv {\bf \Pi}_n \cdot {\bf e}_{\theta}({\bf r}_n)$, and
we have used
\begin{equation}
U^\dagger {\bf \Pi}_n U = {\bf \Pi}_n + {1 \over r_n}
\ \! {\bf e}_{\theta}({\bf r}_n) \ \! \delta \! L .
\end{equation}
The energy difference, $\delta \! E \equiv E' - E$, plotted as a function of
the
parameter $\delta \! L$, is shown in Fig.~1.
For values of $\delta \! L$ such that
\begin{equation}
0 < \delta \! L < {\delta \! L}^*,
\label{inequality}
\end{equation}
where ${\delta \! L}^*$ is the zero of $\delta \! E$ defined in Fig.~1,
the rotating state $\psi'$ has a lower mean energy than $\psi$, so $\psi$
cannot be the ground state.

This is the essential content of Bloch's theorem.
It applies whenever there is a nonzero $\delta \! L$ satisfying
(\ref{inequality}).
However, the smallest nonzero $\delta \! L$ permitted by the condition that the
wave function $\psi'$ be single-valued is $\delta \! L = \hbar$.
When the relativistic corrections in
(\ref{Hamiltonian}) are neglected,
${\delta \! L}^* = {\delta \! L}^*_0$, where
\begin{equation}
{\delta \! L}^*_0 \equiv {4\pi mI\over Ne}\bigg\langle{1\over
r^2}\bigg\rangle^{-1}.
\label{L0}
\end{equation}
Here
\begin{eqnarray}
\bigg\langle{1\over r^\gamma}\bigg\rangle & \equiv &
{1 \over N} \bigg\langle \sum_n {1\over r_n^\gamma}\bigg\rangle
= {1 \over N} \int d^3r \ {n({\bf r}) \over r^\gamma } ,
\end{eqnarray}
where $n({\bf r})$ is the ground-state number density.
Therefore, Bloch's theorem applies only when
${\delta \! L}^* > \hbar$, or whenever
$|I| > I^0_{\rm max}$, where
\begin{equation}
I^0_{\rm max} \equiv {N e \hbar \over 4 \pi m}
\bigg\langle{1\over r^2}\bigg\rangle.
\label{I0}
\end{equation}
This is the upper bound derived in Ref. \cite{Vignale}.
When the relativistic corrections are included to leading order,
${\delta \! L}^*$ is given by
\begin{eqnarray}
{\delta \! L}^* &=& {\delta \! L}^*_0
+ { \pi I \over e m c^2 N^2 \langle 1/r^2 \rangle^2 }
 \bigg\langle \sum_n
\bigg( \Pi_n^2 {1 \over r_n^2}
+ {1 \over r_n^2} \Pi_n^2
+ 4 {(\Pi_n^\theta)^2
 \over r_n^2 } \bigg)
\bigg\rangle \nonumber \\
&+& {2 \pi e I \over c^2 N^2 \langle 1/r^2 \rangle^2 }
\bigg\langle \sum_{n \neq n'}  {
e^i_{\theta}({\bf r}_n) \ T^{ij}({\bf r}_n - {\bf r}_{n'}) \  e^j_{\theta}({\bf
r}_{n'})
\over r_n  | {\bf r}_n - {\bf r}_{n'} | r_{n'} }
\bigg\rangle
+  {16 \pi^2 I^2 \over e^2 c^2 N^3 \langle 1/r^2 \rangle^3 }
\bigg\langle \sum_n { \Pi_n^\theta  \over r_n^3 } \bigg\rangle
+ {16 \pi^3 m I^3 \langle 1/r^4 \rangle \over e^3 c^2 N^3 \langle 1/r^2
\rangle^4 }  .
\label{L}
\end{eqnarray}
Bloch's theorem therefore applies whenever
$|I| > I_{\rm max}$, where
\begin{equation}
I_{\rm max} = (1-\Lambda)  I_{\rm max}^0,
\label{upper bound}
\end{equation}
and
\begin{eqnarray}
\Lambda &\equiv &
{1 \over 4 m^2 c^2 N \langle 1/r^2 \rangle }
 \bigg\langle \sum_n
\bigg( \Pi_n^2 {1 \over r_n^2} + {1 \over r_n^2} \Pi_n^2
+  4 {(\Pi_n^\theta)^2
 \over r_n^2 } \bigg)
 \bigg\rangle
+ {e^2 \over 2 m c^2 N \langle 1/r^2 \rangle }
\bigg\langle \sum_{n \neq n'}
{ e^i_{\theta}({\bf r}_n) \  T^{ij}({\bf r}_n - {\bf r}_{n'}) \
e^j_{\theta}({\bf r}_{n'})
\over r_n  | {\bf r}_n - {\bf r}_{n'} | r_{n'} }
\bigg\rangle  \nonumber \\
&+&  {\hbar \over m^2 c^2 N  \langle 1/r^2 \rangle}
\bigg\langle \sum_n { \Pi_n^\theta \over r_n^3 } \bigg\rangle
+  {\hbar^2 \langle 1/r^4 \rangle \over 4 m^2 c^2  \langle 1/r^2 \rangle}
\label{reduction factor}
\end{eqnarray}
is a dimensionless {\it reduction factor}.
States carrying orbital currents larger than $I_{\rm max}$ cannot be ground
states
of (\ref{Hamiltonian}).

The upper bound (\ref{upper bound}) applies to interacting electron systems in
the
presence of static but otherwise arbitrary electric and magnetic fields,
and includes the effects of spin-orbit coupling and current-current
interaction.
The upper bound (\ref{I0}) applies to noninteracting systems and also to
electrons
with Coulomb interaction. In particular,
(\ref{I0}) applies to noninteracting electrons in a periodic potential, and
this fact
leads to a general constraint on the band structure of any one-dimensional
crystal
\cite{group velocity}.

\section{upper bound for a thin ring}

Now consider the case of a thin ring with cross-sectional dimensions much
less than the radius $R$ of the ring.
In this case,
\begin{equation}
I_{\rm max}^0 = {N e \hbar \over 4 \pi m R^2} = {2 e
v_{\rm \scriptscriptstyle F} \over L},
\label{thin ring I0}
\end{equation}
where
$v_{\rm \scriptscriptstyle F}$
is the Fermi velocity, $L \equiv 2 \pi R$ is the circumference of the ring,
and
\begin{equation}
\Lambda \approx
{3 \over 2 m^2 c^2 N }
 \bigg\langle \sum_n
\big(\Pi_n^\theta \big)^2
 \bigg\rangle
+ {e^2 \over 2 m c^2 N }
\bigg\langle \sum_{n \neq n'}
{ e^i_{\theta}({\bf r}_n) \  T^{ij}({\bf r}_n - {\bf r}_{n'}) \
e^j_{\theta}({\bf r}_{n'})
\over  | {\bf r}_n - {\bf r}_{n'} | }
\bigg\rangle
+  {\hbar \over m^2 c^2 N R }
\bigg\langle \sum_n \Pi_n^\theta \bigg\rangle
+  {\hbar^2 \over 4 m^2 c^2 R^2 }.
\label{thin ring reduction factor}
\end{equation}
The first term in (\ref{thin ring reduction factor}) is approximately equal to
$E_{\rm \scriptscriptstyle F}/mc^2$,
where
$E_{\rm \scriptscriptstyle F}$
is the Fermi energy, and hence this term is entirely negligible here.
The magnitude of the third term may be estimated by using the approximation
$\big\langle \sum_n \Pi_n^\theta \big\rangle \approx 4 \pi m R
I_{\rm max}^0 / e$,
which shows that the third term and fourth term in (\ref{thin ring reduction
factor})
are both of order $\lambda_{\rm c}^2 / R^2$,
where $\lambda_{\rm c} \equiv \hbar/mc$ is the Compton wavelength
of the electron.
These terms are therefore negligible here as well.

The operator in the second term of (\ref{thin ring reduction factor}) may be
written in
second-quantized form as
\begin{equation}
\sum_{n \neq n'}
{ e^i_{\theta}({\bf r}_n) \ T^{ij}({\bf r}_n - {\bf r}_{n'}) \
e^j_{\theta}({\bf r}_{n'})
\over  | {\bf r}_n - {\bf r}_{n'} | }
= \! \int \! d^3r \ \!  d^3r' \ \! F({\bf r},{\bf r'})
\ \! \psi^\dagger({\bf r}) \ \! \psi^\dagger({\bf r}') \ \! \psi({\bf r}') \ \!
\psi({\bf r}),
\label{inductance operator}
\end{equation}
where $\psi({\bf r})$ and $\psi^\dagger({\bf r})$ are electron
field operators, and where
\begin{equation}
F({\bf r},{\bf r}') \equiv
{ e^i_\theta({\bf r})  \ T^{ij}({\bf r} - {\bf r}') \ e^j_\theta({\bf r}')
\over |{\bf r}-{\bf r}'|}.
\label{propagator}
\end{equation}
This term is a consequence of the current-current interactions.
In a mesoscopic or macroscopic system, the largest contribution to the
expectation
value of (\ref{inductance operator}) comes from the direct term
\begin{equation}
\int \! d^3r \ \!  d^3r' \ F({\bf r},{\bf r}') \ n({\bf r}) \ \! n({\bf r}'),
\label{direct term}
\end{equation}
which is normally absent in the case of Coulomb interactions in a uniform
system.
For a thin wire of approximately uniform density and current density, we may
write
the latter as $I$ divided by the cross-sectional area $N/nL$,
\begin{equation}
{\bf j}({\bf r}) \approx { n I L
\over N } \ {\bf e}_{\theta}({\bf r}).
\end{equation}
Then we have
\begin{eqnarray}
 \bigg\langle \sum_{n \neq n'}
{ e^i_{\theta}({\bf r}_n) \ \! T^{ij}({\bf r}_n - {\bf r}_{n'})
\ \!  e^j_{\theta}({\bf r}_{n'})
\over  | {\bf r}_n - {\bf r}_{n'} | } \bigg\rangle
& = & {N^2 \over 4 \pi^2 R^2 I^2}
\! \int \! d^3r \ \!  d^3r'  \ \!
{ j^i({\bf r}) T^{ij}({\bf r}-{\bf r}') j^j({\bf r}') \over
| {\bf r}-{\bf r}'| }        \nonumber \\
& =& {N^2 \over 2 \pi^2 R^2 I^2}
\int d^3r \ d^3r'
\ { {\bf j}({\bf r}) \cdot {\bf j}_{\rm t}({\bf r}') \over
| {\bf r}-{\bf r}' | },
\end{eqnarray}
where ${\bf j}_{\rm t}$ is the {\it transverse} current density, defined as
\begin{eqnarray}
{\bf j}_{\rm t}({\bf r}) & \equiv &
{1 \over 4 \pi} \nabla \times \nabla \times \int d^3r' \ { {\bf j}({\bf r}')
\over |{\bf r}-{\bf r}'| } \nonumber \\
&=& {\bf j}({\bf r}) + {1 \over 4 \pi}
\nabla \int d^3r' { \nabla' \cdot {\bf j}({\bf r}') \over
|{\bf r}-{\bf r}'| }.
\label{transverse current}
\end{eqnarray}
Because the equilibrium current density
satisfies $\nabla \cdot {\bf j} = 0$, the second term
in (\ref{transverse current}) vanishes and ${\bf j}$ is purely transverse.
The reduction factor (\ref{thin ring reduction factor}) for a thin ring may
therefore
be written approximately as
$\Lambda = 2 I_{\rm max}^0 {\cal L} / c \Phi_0, $
where
\begin{equation}
{\cal L} \equiv {1 \over I^2} \int d^3r \ d^3r'
\ { {\bf j}({\bf r}) \cdot {\bf j}({\bf r}') \over
| {\bf r}-{\bf r}'|}
\label{self-inductance}
\end{equation}
is the classical self-inductance of the ring,
and where $\Phi_0 \equiv hc/e$ is the quantum of
flux.
It is also useful to rewrite the reduction factor as
\begin{equation}
\Lambda = {2 I_{\rm max}^0 \over  I_{\rm c}},
\label{inductance reduction factor}
\end{equation}
where
$I_{\rm c} \equiv c \Phi_0 / {\cal L}$ is the
magnitude of the current needed to produce
one quantum of flux.
This latter form makes explicit the relative importance of
the inductive effects. Therefore, the upper bound in a thin ring may be written
as
\begin{equation}
I = \bigg( 1 - { 2 I_{\rm max}^0 \over I_{\rm c} } \bigg) I_{\rm max}^0 .
\label{thin ring upper bound}
\end{equation}
We see that the current-current interactions always
{\it reduce} the allowed ground-state
current by an amount that depends on the self-inductance of the ring.
This reduction occurs because the energy required to sustain a persistent
current now includes the magnetic field energy.
As is clear from our derivation, which treated
the current-current interaction as a small
perturbation, (\ref{thin ring upper bound}) is valid only when $\Lambda << 1$ ,
or when $I_{\rm max}^0 << I_{\rm c}$.
We shall evaluate (\ref{thin ring upper bound}) for realistic thin-ring
geometries
in the final section of this paper.

\section{persistent current in a solvable model with current-current
interactions}

In this section we shall calculate the ground-state current in an electron gas
with current-current interactions, which is confined to a thin wire loop of
circumference $L \equiv 2 \pi R$. We shall assume that only a single transverse
mode in the wire is occupied, so that the system is effectively
one-dimensional,
with a width $a << L$. For modest applied perpendicular magnetic fields, such
that the
magnetic length $\ell \equiv (\hbar c / e B)^{1 \over 2}$ satisfies $\ell >>
a$, the
primary effect of the magnetic field on the electron gas is to induce an
Aharanov-Bohm
phase shift around the ring. The effect of the magnetic field in this
regime may therefore be accounted for by
imposing the twisted boundary conditions
\begin{equation}
\psi_n (x+L) = e^{i 2 \pi \phi} \psi_n(x),
\label{twisted boundary conditions}
\end{equation}
on the single-particle states. Here $x \equiv R \theta$ is an arclength
coordinate going around the circumference of the ring, and
$\phi \equiv \Phi / \Phi_0$
is the total enclosed flux in units of the flux quantum $\Phi_0 \equiv hc/e$.

We shall calculate the ground-state current in a
self-consistent mean field approximation
in which the effect of the current-current interaction is
to increase the flux enclosed by the
ring by an amount that is determined by the
persistent current. The persistent current is,
in turn, determined by the total flux. Coulomb interactions,
which are not expected to affect the persistent
current in a disorder-free system, are ignored altogether.
The single-particle states are
therefore of the form
\begin{equation}
\psi_n = {1 \over \sqrt{L}} \ \! e^{i k_n x},
\label{eigenstates}
\end{equation}
where $n$ is an integer and, according to
(\ref{twisted boundary conditions}), the allowed
wave numbers are
$k_n \equiv 2 \pi (n + \phi)/L$.
The energies of the states (\ref{eigenstates}) are
$\epsilon_n = \hbar^2 k_n^2 / 2 m$,
and
\begin{equation}
I_n = - {2 \pi e \hbar \over m L^2} (n + \phi)
\end{equation}
are their contributions to the total persistent current.

Let $N$ be the total number of electrons, which we shall take to be
twice an odd integer.
(A similar analysis applies to the case of $N/2$ even.)
The total persistent current at zero temperature is given by
\begin{equation}
I = 2 \sum_n \theta(E_{\rm \scriptscriptstyle F} - \epsilon_n) \ \! I_n,
\label{total persistent current}
\end{equation}
where $E_{\rm \scriptscriptstyle F}$ is the Fermi energy,
$\theta(x)$ is the unit step function, and
where the factor of 2 accounts for the spin degeneracy.
For a small flux,
the set of occupied states in (\ref{total persistent current})
is not changed by the flux.
Then we find
\begin{equation}
I = - 2 I_{\rm max}^0 \phi ,
\label{general persistent current}
\end{equation}
which is the well-known result for the persistent current in a one-dimensional
ring
in the presence of an external dimensionless flux $\phi$,
when disorder and electron-electron interactions are ignored.
The expression (\ref{general persistent current}) is
valid until the set of occupied states is changed
by the flux. In the usual case where $\phi$ is independent of $I$, the range of
allowed
flux is $-{\textstyle{1\over 2}} < \phi < {\textstyle{1\over 2}} $, and the
upper bound
(\ref{thin ring I0}) is realized when $\phi \rightarrow \pm {1 \over 2}$.

In our case of interest, however, there are two contributions to $\phi$,
\begin{equation}
\phi = \phi_1 + \phi_2.
\end{equation}
The first,
$\phi_1$, is the flux from the external magnetic field,
and the second, $\phi_2$,
is the flux that
originates from the current-current interactions.
The latter is the flux through the ring induced classically
by the current $I$,
\begin{equation}
\Phi_2 = {1 \over c} \int_{\rm ring} \! \! d{\bf a} \cdot \nabla
\times \int d^3r' { {\bf j}({\bf r'}) \over
|{\bf r}-{\bf r}'| } .
\end{equation}
In a thin ring, this leads to
$\phi_2 = I {\cal L} / c \Phi_0 $,
where ${\cal L}$ is the self-inductance.
Therefore, we obtain from (\ref{general persistent current}),
\begin{equation}
I = -2 I_{\rm max}^0 \ \! \phi_1 \bigg( 1 -
2 {I_{\rm max}^0 {\cal L} \over c \Phi_0 } \bigg) .
\ \ \ \ \ (- {\textstyle{1\over 2}} < \phi_1 < {\textstyle{1\over 2}} )
\label{current for arbitrary applied flux}
\end{equation}
In this simple model, the current is zero when the external flux is zero.
The maximum current
occurs when $\phi_1 \rightarrow \pm {1 \over 2}$.
The magnitude of the persistent current in this optimal case is then found to
be
\begin{equation}
|I| = I_{\rm max}^0 \bigg( 1 -
{ 2 I_{\rm max}^0 {\cal L} \over c \Phi_0 } \bigg),
\label{optimal persistent current}
\end{equation}
which realizes our thin-ring upper bound (\ref{thin ring upper bound}) for
electrons with current-current interactions.

\section{discussion}

We now evaluate the upper bound (\ref{thin ring upper bound}) for realistic
thin-ring
geometries. The upper bound (\ref{thin ring I0}) for a metal with a Fermi
velocity
of $2 \! \times \! 10^{8} \ {\rm cm/s}$, a typical value, may be written as
\begin{equation}
I_{\rm max}^0 \approx { 0.64 \ \! \mu {\rm A} \over L (\mu {\rm m}) },
\label{I0 in amps}
\end{equation}
where $L (\mu {\rm m})$ is the circumference of the ring in microns.
As discussed above, the relevance of inductive effects are characterized by the
ratio of $I_{\rm c} \equiv c \Phi_0 / {\cal L}$, which is the current needed to
produce
one quantum of flux, to $I_{\rm max}^0$.
If we measure the self-inductance in microns, then
\begin{equation}
I_{\rm c} \approx { 41.2 \ \! {\rm mA} \over {\cal L} (\mu {\rm m}) }.
\label{Ic in amps}
\end{equation}
The reduction factor (\ref{inductance reduction factor}) for a metal ring
may then be written as
\begin{equation}
\Lambda \approx 3.0 \! \times \! 10^{-5} \ { {\cal L} \over L} .
\label{metallic reduction factor}
\end{equation}

The self-inductance of a thin toroidal ring with major radius $R$ and minor
radius
$a$ (wire radius) is
${\cal L} = 4 \pi R [ \ln(8R/a) - {\textstyle{7 \over 4}}].$
Therefore, we see that $\Lambda$
depends only on the aspect ratio
$R/a$
of the ring, and not on its circumference:
\begin{equation}
\Lambda \approx 6.0 \! \times \! 10^{-5} \bigg[ \ln \bigg(  {8 R \over a}
\bigg)
-  {7 \over 4} \bigg].
\end{equation}
For a thick ring, where $R/a \approx 1$, ${\cal L}$ is approximately
equal to the size $L$ of the ring. For a thin ring with
$R/a \approx 10$ or $R/a \approx 100$, ${\cal L}$ is substantially larger.

Consider, for example, a gold ring with $L \approx 12 \mu {\rm m}$
and $R/a \approx 30$, characteristic of the rings studied by
Chandresakhar {\it et al.}
\cite{Chandresakhar etal},
where persistent currents of order $I_{\rm max}^0$ where measured.
Then ${\cal L}/L \approx 7.5$, and
$\Lambda \approx 10^{-4}$, a negligible reduction that is consistent with
the experiments.

\acknowledgements

This work was supported by the National Science Foundation through Grant
DMR-9403908. I would like to thank Giovanni Vignale for many useful discussions
on this subject.
I also acknowledge the hospitality of the
Institute for Theoretical Physics,
Santa Barbara, where part of this work was completed under
NSF Grant PHY89-04035.

\begin{figure}
\caption{Energy difference between the ground state $\psi$ and the rotating
state $\psi'$,
as a function of the imparted angular momentum $\delta \! L$.}
\label{figure1}
\end{figure}

\end{document}